\documentstyle[12pt]{article}

\def\[{\left\lbrack}
\def\]{\right\rbrack}

\def\({\left(}
\def\){\right)}
\def\{{\left \{}
\def\{{\right \}}
\def\ih{\'\i}

\title{A NOTE IN THE SKYRME MODEL WITH HIGHER DERIVATIVE TERMS}

\author{Jorge Ananias Neto\\
 Centro Brasileiro de Pesquisas F\ih sicas,\\ R. Dr. Xavier Sigaud 150
22290-180 Rio de Janeiro, Brazil }

\date{February, 1994}

\begin{document}

\maketitle

\begin{abstract}
\noindent Another stabilizer term is used in the classical Hamiltonian of
the Skyrme Model that   permits in a much simple way the generalization
of the higher-order terms in the pion derivative field. Improved
numerical results are obtained.
\end{abstract}

\vskip 1 cm


\newpage

\noindent It is well established that the Skyrme Model\cite{Skyrme}
 reproduces with relative success most of the static properties of
 Nucleon(within approximately
$\sim$ 30\%). The idea consists of treating baryons as soliton solutions in
 the  Non-Linear Chiral SU(2) Sigma Model whose original Lagrangian is
\begin{equation}
\label{l1}
L_1={F^2_\pi\over16} \int d^3r \, Tr\left(\partial_\mu U \partial^\mu U^+
 \) \,,
\end{equation}

\noindent where U is a SU(2) matrix and $\,F_\pi\,$ is
 the pion decay constant. According to Derrick theorem \cite{Derrick}
it is necessary to add to expression (\ref{l1}) an adhoc stabilizer term
\begin{equation}
\label{l2}
L_2={1\over32e^2} \int d^3r \,Tr\[ U^+ \partial_\mu U ,
U^+ \partial_\nu U \]^2 \,,
\end{equation}

\noindent where {\it e} is a dimensionless parameter.

\begin{sloppypar} The physical properties are then calculated making
 use of the semi-classical expansion of the Quantum Hamiltonian
where we perform a rotacional collective coordinate
expansion\cite{ANW} $\,U(r,t)=A(t)U_0(r)A^+(t),\,$ where A is a SU(2)
matrix and $U_0$ a
static soliton solution. $\,$\ Adopting the hedgehog ansatz $
U={\exp i \tau . \hat r F(r)} \,$ where $\tau$ is the Pauli matrix and F(r)
is called the chiral angle,$\,$ the Hamiltonian with the rotational mode can
be written as
\end{sloppypar}

\begin{equation}
\label{HQ}
H= M+{l(l+2)\over 8I}\,, \,\,\, \, l=1,2,\dots ,
\end{equation}
\noindent where $\,M$ is the classical energy and $\,I$ is the inertia
moment\cite{ANW}.

\begin{sloppypar}
\par Dub\'e and Marleau \cite{Marleau} indicate therefore a possible way of
 generalizing the Skyrme Model introducing higher-order terms in the
derivatives  of the pion field.In order to simplify the model, they
 made a particular
 choice of the Hamiltonian parameters that permited  to sum the series in an
exponential funcional form. Using this method they obtained  improved
 numerical results.
\par In the present note we propose to introduce another
 four derivative
stabilizer term \cite{Adkins}given by
\end{sloppypar}

\begin{equation}
\label{l2g}
 L_2 = \int d^3r \, c_2 \[ tr \( \partial_\mu U
\partial^\mu U^+ \) \]^2 \,.
\end{equation}
\noindent As this term is the square of the kinetic term defined in (\ref
{l1}),$\,$ it is natural to take it as a pattern for the inclusion of the
higher-order derivative terms. Thus,$\,$ the standard form of the
Lagrangian terms is
\begin{equation}
\label{ln}
 L_n = \int d^3r \, c_n \[Tr\( \partial_\mu U \partial^\mu U^+ \) \]^n \,,
\end{equation}

\begin{sloppypar}
\noindent where n=1,2,$\dots \,\,\,$. It must be noted
 that this form is also
invariant under chiral  transformation (U transforming under
$U\rightarrow AUB^{-1},\,$ A and B are SU(2) matrices), and does not
destabilize the soliton solution, since we can to define a positive
static Hamiltonian.
\end{sloppypar} To simplify the calculations we will adopt the Sugawara form,
$\, L_\mu = U^+ \partial_\mu U,\,\,$ whose the static component can
be written as $\,\, L_i=i\tau^a L^a_i. \,$ Using the hedgehog
ansatz, the kinetic static term (\ref{l1}) becomes then,

\begin{equation}
L_1 = -c_1 \int d^3r \, Tr \[ \partial_i U \partial_i U^+ \]
 =c_1 \int d^3r \, tr\[ L_iL_i \]
= -2 c_1 \int d^3r \, L^a_iL^a_i \,,
\end{equation}

\noindent where $\,\, L^a_iL^a_i = {2 \sin^2F\over r^2} + F'^2,\,\,$
and $c_1=F^2_{\pi}/16.\,\,$ Since the classical energy is given
by -L, where L is the generalized static Lagrangian, we can write
the classical total positive energy with the inclusion of the
higher order derivative terms as

\begin{eqnarray}
\label{EC}
 & E= \int d^3r \, \[ c_1 \( 2L^a_i L^a_i \) + c_2 \(2L^a_i L^a_i \)^2
+ \dots + c_n \(2L^a_i L^a_i \)^n \] \nonumber \\
&=\int d^3r \,\,  2L^a_iL^a_i \,c_1\[1+
{c_2 \over c_1} \( 2L^a_i L^a_i \)+{c_3\over c_1}\(2L^a_iL^2_i\)^2+\dots+
{c_n \over c_1}{\( 2L^a_i L^a_i \)}^{n-1} \] \,.
\end{eqnarray}

\noindent There are many particular values of the relations
$\,{c_n\over c_1} \,$
in(\ref{EC}) that allow the energy series to summed. Following
Dub\'e and Marleau, if we naively set

\begin{equation}
\label{Kn}
K_n\equiv {c_n\over c_1} ={1\over (n-1)!}\,\,{1\over {(2e^2 F^2_\pi)}^{n-1}}
\,,
\end{equation}

\noindent the series of the classical energy(\ref{EC}) converges in an
exponential form

\begin{eqnarray}
\label{EH}
&E= \int d^3r \, 2c_1 L^a_iL^a_i \exp\[{L^a_iL^a_i\over
e^2F^2_\pi}\] \nonumber \\ & ={F^2_\pi \over 16}
\int d^3r 2\[{2\sin^2F\over r^2}+F'^2\] \exp\[ {{2sin^2F \over r^2}
+F'^2\over e^2F^2_\pi}
 \] \nonumber \\ & =  {F_\pi \over e}\,{\pi\over 2}
 \int^\infty_0 dx x^2 \[ {2sin^2F \over x^2}+F'^2 \]
 \exp\[ {2sin^2F \over x^2}+F'^2 \] \,,
\end{eqnarray}

\noindent with $F_\pi$ and {\it{e}} being the only input parameters.
In expression(\ref{EH}) we have used the series representation $ \,{\exp y}
=\sum^\infty _{k=0} {y^k\over K!} \, $ and a dimensionless variable
x defined by

\begin{equation}
\label{dl}
 x = e F_\pi r \,.
\end{equation}

\noindent It is interesting to point out that with the choice of
standard form(\ref{ln}) and with the definition (\ref{dl}),$\,$ all
the coupling constants are absorbed in the new dimensionless variable x.
{}From (\ref{EH})
the variational equation is

\begin{eqnarray}
\label{Euler}
 &\[2x^2+2{x^2}S+8{x^2}F'^2+4{x^2}SF'^2 \] F^{\prime \prime}
 +8\sin2F F'^2 \nonumber \\
& +4S \sin2F F'^2 + 4xF' - {16sin^2F F'\over x} + 4xSF' \nonumber \\
& -{8Ssin^2F F'\over x} - 2sin2F - 2Ssin2F =0 \,,
\end{eqnarray}

\noindent where $S\equiv \[{2\sin^2F \over x^2}+F'^2 \] $. The soliton
 solution with the baryon number 1 has the boundary
conditions \ $F(x)=\pi$ at $x=0$ and $F(x)=0$
 at $x \rightarrow \infty $. Then, we
impose the asymptotic behaviour of F given by(\ref{Euler})

\begin{equation}
 \lim_{x\rightarrow \infty} F(x) = {B\over x^2} \,,
\end{equation}

\noindent where  B is a constant, to obtain using numerical integration
the soliton solution.$\,\,$ In figure 1, we show the numerical
behaviour of parameter B, which is direclty proportional to the axial
 vector constant coupling,$\,g_A={2\pi\over 3} {B\over e^2}\,$.

\par The inertia moment is calculated similarly to the
series of the classical energy. Performing the rotacional
collective coordinate
expansion of the classical Lagrangian and picking up only terms
linear in $\,Tr(\partial_0A^+ \partial_0A)\,$ ,
we obtain the expression of the series of the inertia moment given
by

\begin{equation}
I =\int d^3r {8\over3} \,\sin^2F \,\,c_1 \[ 1+2K_2 \( 2L^a_i L^a_i \)
 +\dots +nK_n \( 2L^a_i L^a_i \)^{n-1} \] \,.
\end{equation}

\noindent With the substitution of  $K_n$ by the expression(\ref{Kn}), using
the dimensionless variable {\it x} defined in(\ref{dl}) and taking the limit
$n\rightarrow \infty, \,\,$ we find

\newpage

\begin{eqnarray}
I={2\pi\over3} \,{1\over F_{\pi} e^3}\int^\infty_0 dx x^2
\sin^2F \[ 1+F'^2+{2\sin^2F\over x^2}\] \nonumber \\
\hskip 3 cm \exp\[ F'^2+{2\sin^2F \over x^2} \] \,,
\end{eqnarray}
\noindent where we have used the series representation $\, {\exp y}(1+y)
= \sum^\infty_{k=0} {y^k(k+1) \over k!} \,$.
\begin{sloppypar}\noindent  The numerical calculations are then performed
 using the masses
of the Nucleon   $ (M_N=939 Mev)$ and of the Delta $ (M_\bigtriangleup=
1232 Mev)$ as input parameters to determine the pion decay constant
 $ F_ \pi $
and the dimensionless Skyrme parameter $ {\it e}.\,\,$\noindent
The main physical results are shown in table 1,
 according to\cite{ANW}.\end{sloppypar}
\vskip .8 cm

%\begin{figure}
%\end{figure}

\vskip 10cm
\noindent Fig 1. Behaviour of the parameter B definided by
$B\equiv x^2F(x)\,,$ where F(x) is the numerical variational
solution of the classical Hamiltonian including terms up to
order 4(n=2), 6(n=3), 8(n=4) (solid line) and all orders
(dashed line) in derivatives of pion field.

\newpage
\centerline{TABLE 1- Physical parameters in the Skyrme Model}

$$\vbox{\halign{\hfil#\hfil&\qquad\hfil#\hfil&\qquad\hfil#\hfil&
\qquad\hfil#\hfil&\qquad\hfil#\hfil\cr
\noalign{\hrule}
\ \cr
 &Adkins\cite{ANW} & Marleau\cite{Marleau} & This model & Expt. \cr
\ \cr
\noalign{\hrule}
\ \cr

{F$_\pi(Mev)$} & 129 & 146 & 144 & 186 \cr

{\it e} & 5.45 & 8.69 & 6.82 & - \cr\cr

$<r^2>^{1\over2}_{I=0}(fm)$ & 0.59 & 0.60 & 0.61 & 0.72\cr

$\mu_p$ & 1.87 &  1.89 & 1.90 & 2.79\cr

$\mu_n$ & -1.33 & -1.32 & -1.31 & -1.91\cr

$g_A$ & 0.61 & 0.71 & 0.80 & 1.23\cr

\ \cr
\noalign{\hrule}}}$$

\vskip 2 cm

\noindent Our results indicate an improvement in the physical values of the
 pion decay constant,$\,F_\pi\,$, and the  axial coupling constant,
$ g_A $. The others values, i.e., the  magnetic moments and the
isoescalar mean square radius remain basicaly the same obtained
by Adkins\cite{Adkins} et al. and Marleau\cite{Marleau}.
This procedure, without doubt, improves the physical results.
 In order to obtain  better results for quantities like  magnetic moments and
the  isoescalar mean square radius, we should  treat
the quantization of the classical Hamiltonian
(collective coordinate quantization \cite{ANW}) in more formal way, using
the information about constraint that is  present in the system\cite{EU}.
This particular study will be object of a forthcoming paper \cite{EU2}.
\thanks{

\par I would like to thank M.G. do Amaral for critical reading. The work is
supported by CNPq,
 Brasilian Research Council. }

%\newpage


\begin{thebibliography} {99}


\bibitem{Skyrme}
T.H.Skyrme, Proc. Roy. Soc. A260, 127(1961).

\bibitem{Derrick}
G.Derrick, J. Math. Phys. 5, 1252 (1964).

\bibitem{ANW}
G.S.Adkins,C.R. Nappi and E. Witten, Nucl.Phys. B228,553 (1983).

\bibitem{Marleau}
S.Dub\'e and L.Marleau, Phys.Rev.D41,5,1606 (1990).

\bibitem{Adkins}
G.S.Adkins, Chiral Solitons, Proceedings of the Lewes Workshop
(World Scientific, Singapore) pag.47.

\bibitem{EU}
K.Fujii, K.I.Sato, N.Toyota and A.P. Kobushkin, Phys. Rev. Lett.58,7 (1987),
Jorge Ananias Neto, Preprint IF/UFRJ/94, HEP-TH 9401001.

\bibitem{EU2}
Jorge Ananias Neto, work in progress.

\end{thebibliography}
\end{document}